\documentclass[10pt]{iopart}
\usepackage{graphicx}
\begin{document}

\title{Free-carrier dynamics in Au$_{2}$Pb probed by optical conductivity measurements}


\author{R Kemmler$^{1}$, R H\"ubner$^{1,2}$, A L\"ohle$^{1}$, D Neubauer$^{1}$,
\\ I Voloshenko$^{1}$, L M Schoop$^{3}$, M Dressel$^{1}$ and A V
Pronin$^{1}$}

\address{$^{1}$ 1. Physikalisches Institut, Universit\"at Stuttgart,
70569 Stuttgart, Germany}

\address{$^{2}$ Biomedical Chemistry, Department of Clinical Radiology and Nuclear Medicine,
Medical Faculty Mannheim of Heidelberg University, 68167 Mannheim,
Germany}

\address{$^{3}$ Department of Chemistry, Princeton University, Princeton, NJ 08544, USA}


\eads{artem.pronin@pi1.physik.uni-stuttgart.de}

\vspace{10pt}
\begin{indented}
\item October 25, 2018
\end{indented}

\begin{abstract}
We measured the optical reflectivity of the Dirac material
Au$_{2}$Pb in a broad frequency range (30 -- 48\,000~cm$^{-1}$) for
temperatures between 9 and 300~K. The optical conductivity, computed
from the reflectivity, is dominated by free-carrier contributions
from topologically trivial bulk bands at all temperatures. The
temperature-independent total plasma frequency of these carriers is
$3.9 \pm 0.2$ eV. Overall, optical response of Au$_{2}$Pb is
typically metallic with no signs of localization and bad-metal
behavior.
\end{abstract}


\section{Introduction}

Current tremendous interest in different topological materials has
led to prediction and experimental verification of a number of new
solid-state phases, where the low-energy physics is governed by
novel emerging quasiparticles, such as Dirac or Weyl
fermions~\cite{Wehling2014}. Recently, Au$_{2}$Pb gained attention
because of the prediction~\cite{Schoop2015} of Dirac bands in this
Laves-phase compound, which was known to be superconducting ($T_{c}
\simeq 1.2$ K) for decades \cite{Hamilton1965}. Au$_{2}$Pb possesses
three structural transitions at $T = 97$, 51 and 40 K
\cite{Schoop2015, Chen2016, Yu2016, Xing2016, Wu2018}. The
high-temperature structure is cubic (space group $Fd\bar{3}m$) and
the electronic structure features bulk Dirac cones (six per
Brillouin zone). The Dirac points are protected by $C_{4}$ rotation
symmetry. Upon cooling, the symmetry gets lowered and the Dirac
bands become gapped. At the lowest temperatures, the structure is
orthorhombic (space group $Pbcn$). This phase has also been
predicted to be topologically nontrivial and to possess topological
surface states with linear dispersion relations~\cite{Schoop2015}.
In the superconducting state, the surfaces of Au$_{2}$Pb crystals
are therefore a natural platform for realizing Majorana
fermions~\cite{Fu2008}. Most recent angle-resolved photoemission
spectroscopy (ARPES) provided an experimental evidence for Dirac
states in Au$_{2}$Pb~\cite{Wu2018}.

A direct gap is present in the low-temperature electronic structure
at all momenta, but a number of electron- and hole-like trivial
bands cross the Fermi level. The trivial electron and hole pockets
also exist in the high-temperature phases, in addition to the bulk
Dirac bands. Carriers from these trivial Fermi-surface pockets
dominate the broad optical response of Au$_{2}$Pb. In this paper, we
investigate their contributions to the optical conductivity.

\begin{figure}[b]
\centering
\includegraphics[width=10 cm,clip]{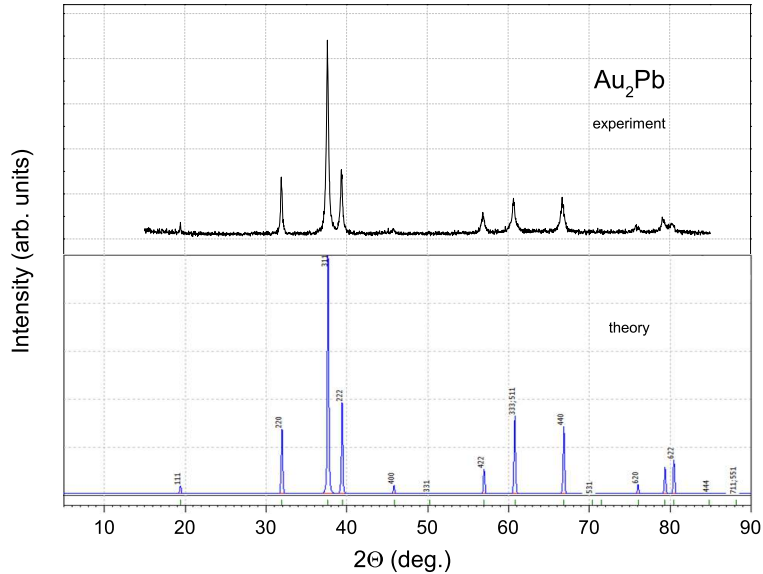}
\caption{X-ray diffraction pattern of milled Au$_{2}$Pb powder (top
panel) in comparison with calculations (bottom panel).} \label{xrd}
\end{figure}

\section{Sample preparation and characterization}

Single crystals of Au$_{2}$Pb were synthesized according to the
procedure reported by Schoop \textit{et al.}~\cite{Schoop2015}. The
single crystals were grow out of lead flux. For the synthesis, 0.78
g of gold powder (99.99 \%) and 1.24 g of lead beads (99.999 \%,
average diameter is 1 mm) were put together in a quartz tube with a
neck of about 1-2 mm in the middle of the tube. The evacuated and
sealed tube was heated within 12~h to 600~$^{\circ}$C and kept there
for 24~h. Then the melt was allowed to cool down (3~$^{\circ}$C/h)
to 300~$^{\circ}$C and kept there for 48~h. When the quartz glass
tube was turned, flux and crystals were separated. The traces of
remaining lead on the surface of the crystals were removed by
washing them in an aqueous solution of 20~ml acetic acid (50~\%) and
4~ml hydrogen peroxide (30~\%) for some minutes. Powder x-ray
diffraction data of ball milled samples were collected on a Bruker
D2 Phaser using MoK$\alpha$-radiation ($\lambda$ = 0.71073 {\AA}) at
room temperature and indicate a cubic Laves phase (space group
$Fd\bar{3}m$) with the lattice constant $a = 7.911(1)$ \AA. No
reflections from impurities, elements (Au and Pb), and any other
lead-gold phases have been detected in these measurements, see
figure~\ref{xrd}. For the crystals used in the measurements, Laue
diffraction demonstrated perfect single-crystalline patterns with
the largest surfaces being either (111) or (100).

\begin{figure}[t]
\centering
\includegraphics[width=\columnwidth, clip]{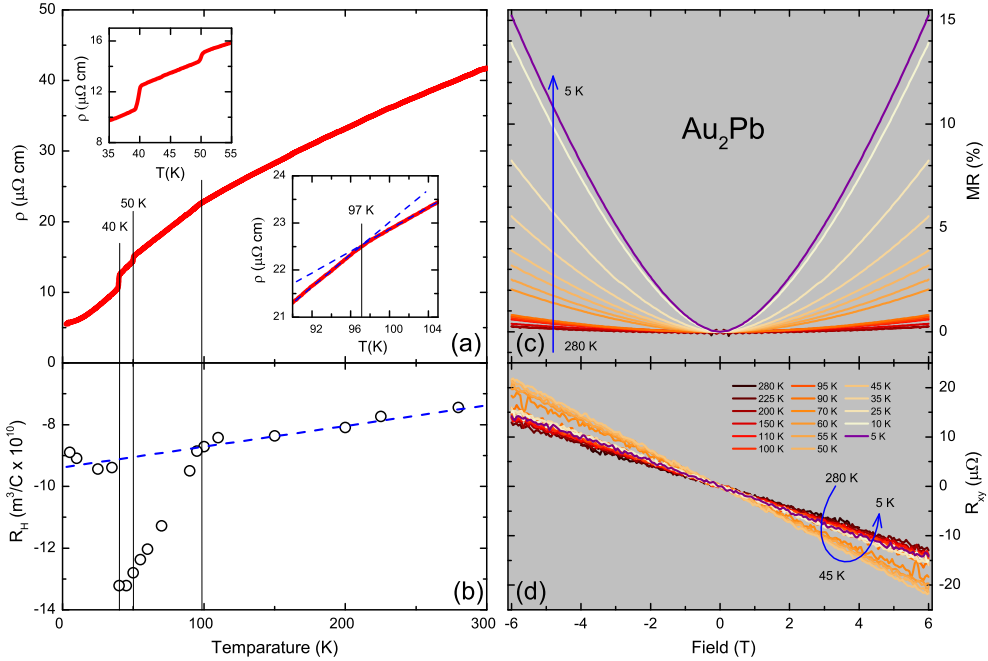}
\caption{Panel (a): Electrical dc resistivity of Au$_{2}$Pb as a
function of temperature. The insets zoom in the areas near the
structural transitions at 40, $\sim$ 50 and 97 K. Panel (b): Hall
coefficient as a function of temperature (open dots). The structural
transitions are indicted as thin vertical lines in the panels (a)
and (b). In both panels, dashed blue lines serve as guides to the
eye. Panel (c): Transversal magnetoresistance (in percent) as a
function of applied magnetic field for a number of temperatures
between 5 and 280 K. Panel (d): Hall resistance as a function of
magnetic field for the temperatures indicated.} \label{dc}
\end{figure}

Electrical dc resistivity measurements were performed in
four-contact geometry in a custom-made setup by cooling from room
temperature down to 5~K. Transversal magnetoresistance (MR) and Hall
resistivity measurements were conducted on Hall bars at 5 to 280 K
in magnetic fields $B$ of up to 6~T. The crystal plane used for all
transport experiments was (110). We should notice that due to the
cubic high-temperature structure and the fact the at low
temperatures the orthorhombic distortions are pretty
small~\cite{Schoop2015}, no significant anisotropy is expected in
Au$_{2}$Pb.

The results of our dc measurements are shown in the panel (a) of
figure~\ref{dc}. A clear metallic behavior is observed, the residual
resistivity being $\rho_0$ = 5.5~$\mu\Omega$cm and $\rho$($300$
K)/$\rho_0 = 7.5$. These values, as well as the overall resistivity
curve, are in good agreement with results previously reported in
literature~\cite{Schoop2015, Chen2016, Yu2016}. As discussed above,
Au$_{2}$Pb undergoes several structural phase transitions with
decreasing temperature: at 97, 51 and 40 K~\cite{Schoop2015};
accordingly, anomalies in the measured $\rho(T)$ are detected at
these temperatures. The magnetotransport results are presented in
the panels (b), (c) and (d). We observe a non-saturating moderate
positive magnetoresistance at all measurement temperatures and
magnetic fields, panel (c). The Hall constant is negative and
non-monotonic in temperature [panels (b) and (d)]. It exhibits jumps
at the structural-transition temperatures, reflecting the
corresponding changes in band structure~\cite{Schoop2015}.

As already noticed, band-structure calculations for the high- and
low-temperature phases predict that both, electron- and hole-like,
bands cross the Fermi level and, hence, contribute (with different
signs) to $R_{H}$ at any temperature. Thus, a straightforward
determination of the free-carrier concentration from $R_{H}$ is not
possible. Application of the one-carrier-type (OCT) relation,
$R_{H}=1/(en)$, allows a low-bound estimation for the free-carrier
concentration, $n \sim 5 \times 10^{21}$ cm$^{-3}$ at 5 K. Such high
$n$ strongly indicates conduction of metallic type (obviously, the
real carrier concentrations can only be higher, as holes and
electrons provide Hall constants of opposite signs, which partly
compensate each other). Within the OCT model, one can also obtain
rough estimates of the carrier mobility, $\mu^{OCT} (5 \rm{K}) \sim
10^{2}$ cm$^{2}$/(Vs), and of the relaxation rate, $\gamma^{OCT} (5
\rm{K}) \sim 100$ cm$^{-1}$ (here, the free-electron mass is taken
as the effective carrier mass). These values are obviously some sort
of averages over all the bands crossing the Fermi level and hence
should be considered with a grain of salt. Still, $\gamma^{OCT}$
lies right in-between of the scattering rates obtained from our
optical studies for the carriers described by the so-called narrow
and broad Drude components, as discussed below.

\section{Optical experiments}

Temperature-dependent optical reflectivity, $R(\nu)$, was measured
on a large [$3 \times 2$ mm$^{2}$] polished (100) surface of a
Au$_{2}$Pb single crystal over a broad frequency range from $\nu =
\omega/(2 \pi c)= 30$ to 10\,000 cm$^{-1}$. Additionally, we
measured room-temperature reflectivity up to 48\,000 cm$^{-1}$. The
spectra in the far-infrared (below 700 cm$^{-1}$) were collected
with a Bruker IFS 113v Fourier-transform spectrometer using
\textit{in situ} gold coating of the sample surface for reference
measurements. At higher frequencies (up to 22\,000 cm$^{-1}$), a
Bruker Hyperion infrared microscope attached to a Bruker Vertex 80v
FTIR spectrometer was used. For these measurements, freshly
evaporated gold mirrors (below 10\,000 cm$^{-1}$) and protected
silver (above 10\,000 cm$^{-1}$) served as reference. Finally, at
the highest frequencies (up 48\,000 cm$^{-1}$) we used a Woollam
variable-angle spectroscopic ellipsometer with SiO$_{2}$ on Si
substrate as reference.

For Kramers-Kronig analysis~\cite{Dressel2002}, zero-frequency
extrapolations have been made using the Drude model in accordance
with the temperature-dependent dc resistivity measurements. Two
narrow gaps in the measurements at around 100 and 700 cm$^{-1}$,
originating from absorption in beam splitters, were bridged by
linear interpolations. For high-frequency extrapolations, we
utilized the x-ray atomic scattering functions~\cite{Tanner2015}
followed by the free-electron behavior, $R(\omega) \propto
1/\omega^{4}$, above 30 keV.

\begin{figure}[t]
\centering
\includegraphics[width=12 cm,clip]{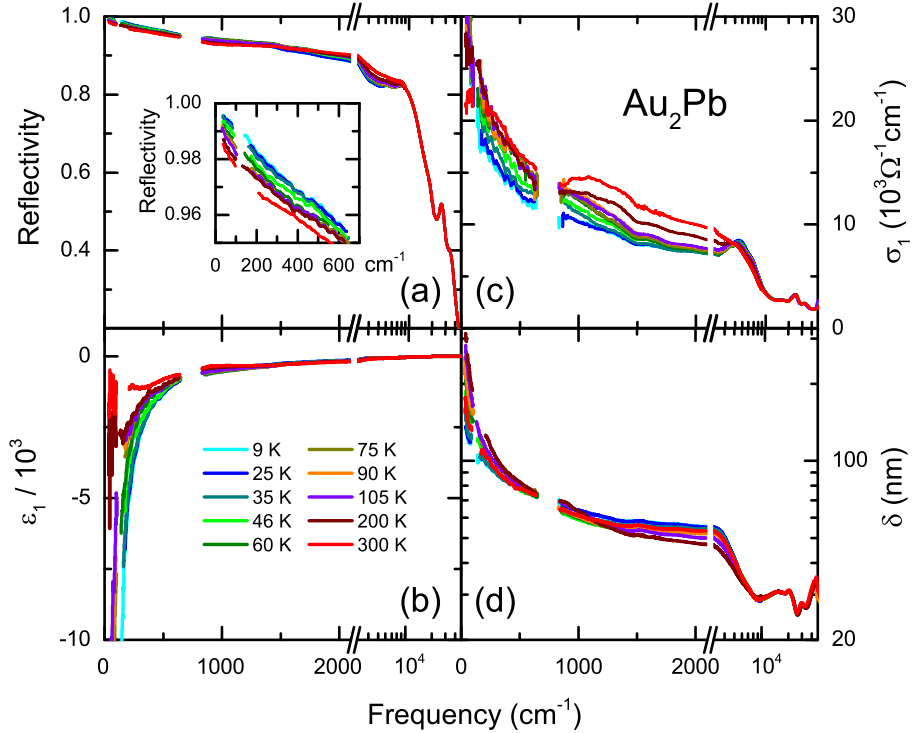}
\caption{Optical reflectivity (a), real parts of the dielectric
permittivity (b) and optical conductivity (c), and skin depth (d) of
Au$_{2}$Pb at selected temperatures between $T=9$ and 300~K; note
the x-scale change at 2000 cm$^{-1}$. The inset shows low-frequency
reflectivity on enlarged scale.} \label{ref_sig_eps}
\end{figure}

\begin{figure}[t]
\centering
\includegraphics[width=12 cm,clip]{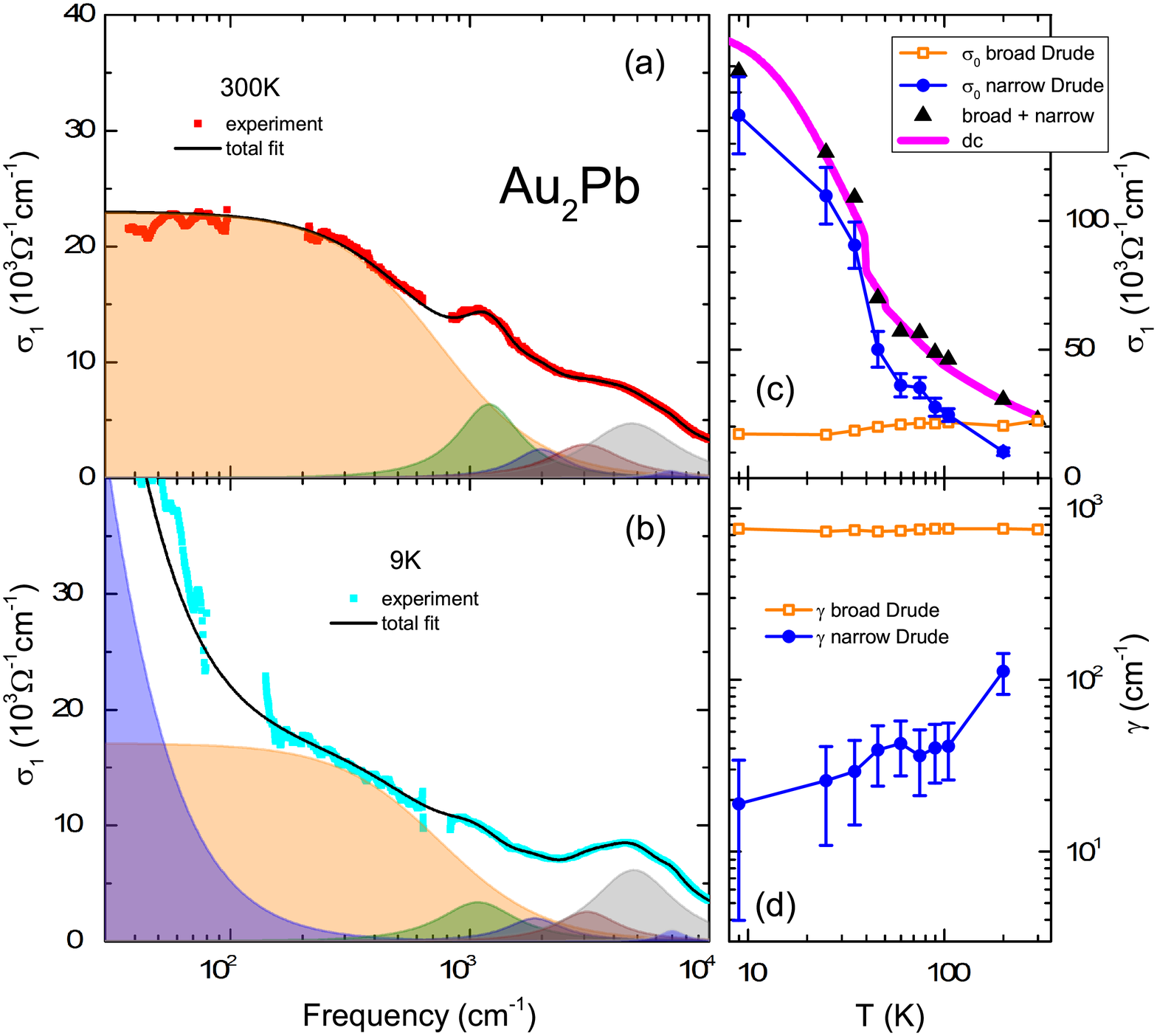}
\caption{Drude-Lorentz fits of the optical conductivity of
Au$_{2}$Pb at 300~K (a) and 9~K (b). Shadowed areas of different
colors correspond to Lorentzians and the two Drude terms (broad and
narrow) as discussed in the text. Parameters of the Drude terms as
functions of temperature: the zero-frequency limit $\sigma_{0}$ (c)
and the scattering rate $\gamma$ (d). Typical fit error bars for
$\sigma_{0}$ and $\gamma$ of the narrow Drude term are displayed;
the error bars for the broad Drude term are within the symbols. Bold
magenta line in panel (c) shows the dc conductivity.} \label{fit}
\end{figure}

Figure~\ref{ref_sig_eps} displays the optical reflectivity $R(\nu)$,
the real part of the optical conductivity $\sigma_{1}(\nu)$ and of
the dielectric constant $\varepsilon_{1}(\nu)$, as well as the skin
depth $\delta(\nu)$ of Au$_{2}$Pb for different temperatures. For
frequencies higher than $\sim$ 10\,000~cm$^{-1}$, the optical
properties are independent of temperature. Let us immediately note
that the skin depth exceeds 20 nm for all measured temperatures and
frequencies. Hence, our optical measurements of Au$_{2}$Pb reflect
its bulk properties.

Au$_{2}$Pb demonstrates a typical metallic response: at $\nu
\rightarrow 0$ the reflectivity approaches unity, $\varepsilon(\nu)$
is large and negative, $\sigma_{1}(\nu)$ is very high, and
$\delta(\nu)$ is inversely proportional to frequency. $R(\nu)$
starts to drop down at around 10\,000~cm$^{-1}$, indicating the
onset of the plasma edge. This high-frequency position of the plasma
frequency is another indication of the high free-carrier
concentration in Au$_{2}$Pb.

Further, we can conclude that high concentration of free carriers
prevents us from observing any sort of characteristic optical
features of a Dirac material, where the interband transitions
between the linearly dispersing bands are supposed to manifest
themselves as a linear-in-frequency $\sigma_{1}(\nu)$ for 3D
systems~\cite{Hosur2012, Bacsi2013, Timusk2013, Chen2015,
Neubauer2016} or as a frequency-independent conductivity in the 2D
case~\cite{Ando2002, Kuzmenko2008, Mak2008, Carbotte2017, Ahn2017,
Schilling2017Zr}.

In Au$_{2}$Pb, one could expect to see the optical transitions
between the lower and the upper Dirac bands at frequencies below
some 3\,200 cm$^{-1}$, as the Lifshitz transition is supposed to be
at around 400 meV $\cong$ 3\,200 cm$^{-1}$~\cite{Schoop2015}. To
estimate the expected interband optical conductivity due to the
inter-Dirac-band transitions $\sigma_{1}^{\rm{Dirac}}$, we can
utilize the well-known relation~\cite{Hosur2012, Bacsi2013}:
\begin{equation}
\sigma_{1}^{\rm{Dirac}}(\omega) = \frac{N_{D} e^2} {6 h}
\frac{\omega} {v_F}, \label{sigma_Dirac}
\end{equation}
which connects the real part of the complex conductivity to the
number of 3D Dirac bands per Brillouin zone $N_{D}$ and the
Dirac-band Fermi velocity $v_F$. Using $N_{D} = 6$ and $v_F \sim
10^6$ m/s, estimated from the band-structure
calculations~\cite{Schoop2015}, we calculate the expected optical
conductivity due to the transitions between the lower and the upper
Dirac bands. We find this conductivity to be orders of magnitude
below the measured values. For example, we obtain
$\sigma_{1}^{\rm{Dirac}}$ of the order of 100 $\Omega^{-1}$cm$^{-1}$
at 1000 cm$^{-1}$ (cf. the experimentally observed value,
$\sigma_{1} \sim 10^{4}$ $\Omega^{-1}$cm$^{-1}$). Hence, the
inter-Dirac-band transitions are completely masked in Au$_{2}$Pb by
free carriers. Hereafter, we concentrate on the intra-band, i.e.
free-carrier, optical response.

In order to fit the optical response of Au$_{2}$Pb, we used a
standard Drude-Lorentz ansatz~\cite{Dressel2002}. We have found that
best description can be obtained, if we utilize two Drude
contributions. Being expressed in terms of complex optical
conductivity, $\hat{\sigma} \equiv \sigma_{1} + \rm{i}\sigma_{2} =
\sigma_{1} + \rm{i}(1-\varepsilon_{1})\omega/(4\pi) $, our fitting
function reads as:
\begin{equation}
\hat{\sigma}= \frac{\sigma^{N}_{0}}{1-\rm{i}\omega\tau^{N}} +
\frac{\sigma^{B}_{0}}{1-\rm{i}\omega\tau^{B}} +
\frac{\omega}{4\pi}\sum
\frac{\Delta\varepsilon_{k}\omega_{0k}^{2}}{\rm{i}(\omega_{0k}^{2} -
\omega^{2} - \rm{i}\omega\Gamma_{k})}. \label{Drude}
\end{equation}
Here, $\sigma_{0}$ is the dc limit of a Drude term, $\tau$
represents a free-electron scattering time, $\omega_{0k}$ is the
eigenfrequency of $k$-th Lorentz oscillator, $\Delta\varepsilon_{k}$
is its dielectric contribution and $\Gamma_{k}$ is the corresponding
line width.

As mentioned above, band-structure calculations and ARPES
measurements demonstrate that at any temperature Au$_{2}$Pb
possesses a few bands (Dirac or not) crossing the Fermi
level~\cite{Schoop2015, Wu2018}. Thus, a multi-component Drude fit
is relevant. Numerous previous optical-conductivity studies show
that two Drude components is typically sufficient to fit the
spectra, even if the system under study possesses more than two
bands: first, adding further Drude terms ($>2$) leads to ambiguous
fits; second, a given Drude term is not necessarily related to the
scattering processes within a given conduction band. Thus, we stick
to the common minimal model. Such two-component approach has been
widely used to interpret the normal-state optical spectra of
iron-based superconductors~\cite{Wu2010, Neubauer2017}. Recently, it
has also been applied to (topological)
semimetals~\cite{Schilling2017Yb, Neubauer2018}. In this case, the
two Drude terms can be interpreted as the response of free carriers
within the Dirac (highly mobile carriers) and parabolic (less mobile
carriers) bands. We show below that this simple interpretation is
not applicable to Au$_{2}$Pb.

Figures~\ref{fit}~(a) and (b) show the results of fitting the
experimental data with Eq.~\ref{Drude} at 9 and 300 K. Let us note,
that for best possible model description of the spectra, we fitted
the experimental spectra of $R(\nu)$, $\sigma(\nu)$ and
$\varepsilon(\nu)$ simultaneously. Also, we did not impose any
restrictions on the fit parameters. In addition to the Drude terms,
we used five Lorentzians, which effectively describe multiple
transitions between trivial bands at high frequencies. We found that
the two-Drude approach is relevant at all measurement temperatures,
except of 300 K, where the two Drude components are not
distinguishable and the spectra can be described by a single Drude
term.

\begin{figure}[t]
\centering
\includegraphics[width=6 cm,clip]{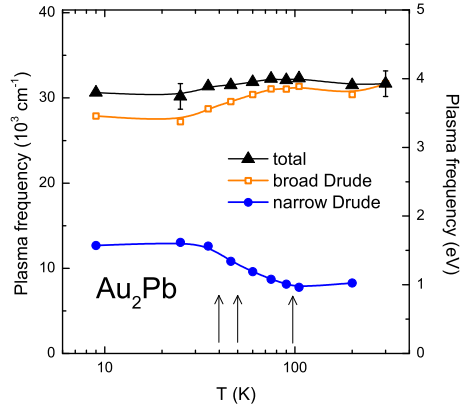}
\caption{Plasma frequencies of the broad and narrow Drude components
and total plasma frequency of free carriers in Au$_{2}$Pb as
functions of temperature. The structural transitions (see
figure~\ref{dc}) are marked with arrows.} \label{plasma}
\end{figure}

The fit parameters of the Drude terms versus temperature are shown
in figures~\ref{fit}~(c) and (d). As one can see from the panel (c),
the two Drude components completely account for the dc conductivity,
$\sigma_{dc}(T) \equiv 1/\rho(T)$, of Au$_{2}$Pb: the sum of the
zero-frequency limits of the Drude modes matches the dc conductivity
values perfectly at any temperature. It is also evident that it is
the narrow component, which is mostly responsible for the
temperature dependence of $\sigma_{dc}$. Overall, the temperature
evolution of the optical spectra is pretty smooth.

We have to notice that the broad Drude term does reflect behavior of
free carriers, has nothing to do with localization and cannot be
replaced by a Lorentzian. As one can see from figure~\ref{fit}~(d),
this term has an almost temperature-independent scattering rate,
$\gamma^{B} = 1/(2 \pi c \tau^{B}) = 750$ cm$^{-1}$. An averaged
Fermi velocity of the non-Dirac bands in Au$_{2}$Pb can be estimated
from the band structure~\cite{Schoop2015} and is in the $10^5$ m/s
range (for Dirac bands, it is even higher, as discussed above).
Thus, the mean free path of the carriers, responsible for the broad
Drude term, is at least around 7 \AA. This is a few times more than
the distance between the atoms. Hence, unlike the situation, e.g.,
in the superconducting cuprates~\cite{Hussey2004, Basov2005} and
some other materials~\cite{Kostic1998, Takenaka1999, Lobo2015}, the
Ioffe-Regel limit~\cite{Ioffe1960} is not violated in Au$_{2}$Pb. It
is not a bad metal~\cite{Emery1995}, the transport in Au$_{2}$Pb is
coherent and our ``broad Drude'' description is relevant.

The plasma frequencies of the two Drude components, $\nu_{pl}^{B}$
and $\nu_{pl}^{N}$, are shown in figure~\ref{plasma}. They have
generally rather flat temperature dependencies. Nevertheless, the
plasma frequency (and hence, the corresponding spectral weight,
which is proportional to the carrier density) of the narrow Drude
band increases upon cooling. Would this absorption band correspond
to the bulk Dirac carriers, $\nu_{pl}^{N} (T)$ had to decrease
instead and eventually to disappear as $T \rightarrow 0$, because
the bulk Dirac bands are gapped and do not cross the Fermi level in
the low-temperature phase of Au$_{2}$Pb. As this decrease of
$\nu_{pl}^{N} (T)$ does not happen, one can conclude that the narrow
Drude term is not directly related to Dirac electrons. Obviously,
the bulk Dirac electrons cannot provide any significant contribution
to the broad Drude term, as this term is present at all temperatures
and its spectral weight has hardly any temperature dependence. Also,
Dirac fermions are typically highly mobile~\cite{Liang2015,
Shekhar2015}, their scattering rates are low and the corresponding
Drude bands are very narrow~\cite{Neubauer2016, Neubauer2018}.
Overall, the contribution of free Dirac carriers to the total
optical response is basically negligible and optics only probes the
carriers in the trivial bands of Au$_{2}$Pb for all its structural
phases.

In figure~\ref{plasma}, the temperatures of the structural
transitions are marked with vertical arrows. It is apparent that at
temperatures around these transitions, the spectral weight
redistributes between the Drude components, reflecting changes in
the band structure. One has to note that plasma frequency is
determinate by both, free-carrier concentration and effective mass.
Thus, sharp changes in the Hall coefficient need not be directly
reflected in the temperature evolution of plasma frequency. The
smooth temperature evolution of $\nu_{pl}^{B}$ and $\nu_{pl}^{N}$
between roughly 40 and 100 K (at higher and lower temperatures, they
are temperature independent within our accuracy) might indicate that
more than one crystallographic structure exists at a given
temperature here. Such nonhomogeneous states have indeed been
detected in Au$_{2}$Pb by temperature-dependent x-ray diffraction at
a number of temperatures in this intermediate-temperature
range~\cite{Schoop2015}.

As one can see from figure~\ref{plasma}, the total plasma frequency
is almost temperature independent, in agreement with the metallic
type of conduction, see figure~\ref{dc}. The temperature-averaged
value of the total plasma frequency is $\nu_{pl} = \sqrt{
(\nu_{pl}^{B})^{2} + (\nu_{pl}^{N})^{2}} = (31\,500 \pm 1\,500)$
cm$^{-1}$, $h\nu_{pl} \cong (3.9 \pm 0.2)$ eV. From $\nu_{pl}$, the
$n/m^{*}$ ratio can be directly computed via:
\begin{equation}
\omega_{pl}^{2} = (2 \pi c \nu_{pl})^{2} = \frac{4\pi e^2 n}
{m^{*}}. \label{plasma_freq}
\end{equation}
Here $m^{*}$ is an averaged effective mass of all carriers and a
parabolic band dispersion is assumed (as discussed above, we can
ignore Dirac electrons, for which the plasma frequency is supposed
to be related to the Fermi velocity, the band filling and,
generally, the band gap~\cite{DasSarma2009, Sachdeva2015}). Thus, we
obtain from equation (\ref{plasma_freq}): $\frac{n}{m^{*}/m_{e}} =
1.1 \times 10^{22} \textrm{cm}^{-3}$, where $m_{e}$ is the
free-electron mass. The average effective mass is not know for
Au$_{2}$Pb. If we take $m^{*} = m_{e}$, we get $n \sim 10^{22}$
cm$^{-3}$, a reasonable metallic concentration.

\section{Summary}

We have measured broadband optical response of Au$_{2}$Pb -- a Dirac
material, which possess bulk Dirac bands at high temperatures (above
97 K) and surface Dirac bands at low temperatures (below 40 K).
Neither of the Dirac states could be seen in optical conductivity
because of the very high absorption due to free carriers (electrons
and holes) in the trivial bulk bands. The total plasma frequency of
these carriers remains temperature independent at $3.9 \pm 0.2$ eV.
Optical measurements provide an estimate for the total carrier
concentration: $\frac{n}{m^{*}/m_{e}} \simeq 10^{22}$ cm$^{-3}$.

\ack We thank Gabriele Untereiner, Annette Zechmeister, Marian
Blankenhorn, Roland R\"osslhuber and Lucky Z. Maulana for valuable
technical support. This work was funded by the Deutsche
Forschungsgesellschaft (DFG) via grant No. DR228/51-1.

\end{document}